\documentclass{PoS}

\title{Tools for Dark Matter in Particle and Astroparticle Physics}
\ShortTitle{Tools for Dark Matter  in Particle and Astroparticle Physics}

\author{\speaker{Alexander Pukhov}\\
       Skobeltsyn Inst. of Nuclear Physics, Moscow State Univ., Moscow, Russia\\
       E-mail: \email{pukhov@lapp.in2p3.fr}}

\author{Genevi\`eve B\'elanger\\
        LAPTH, Univ. de Savoie, CNRS, B.P.110,  F-74941 Annecy-le-Vieux,France\\
        E-mail: \email{belanger@lapp.in2p3.fr}}

\author{Fawzi Boudjema\\
        LAPTH, Univ. de Savoie, CNRS, B.P.110,  F-74941 Annecy-le-Vieux, France\\
        E-mail: \email{boudjema@lapp.in2p3.fr}}

\author{Andrei Semenov\\
        Joint Institute for Nuclear Research (JINR) 141980, Dubna,  Russia\\
        E-mail: \email{semenov@lapp.in2p3.fr}}

\abstract{      Despite   
several  indirect  confirmations of the  existence of dark matter, 
the  properties of a new dark matter particle are still largely unknown.
Several experiments  are currently searching for this particle underground in
direct detection, in
space and on earth in indirect detection and at the LHC. A confirmed signal
could select a model for dark matter among the many
 extensions of the standard model.  In this paper we present  a
short  review    of the public codes for computation of dark matter observables.  
}

\FullConference{13th International Workshop on Advanced Computing and
Analysis Techniques in Physics Research\\
February 22-27, 2010\\
Jaipur, India}

\begin{document}

\def\micro{{\tt micrOMEGAs}}
\def\micromegas{{\tt micrOMEGAs}}
\def\ra{\rightarrow}
\def\calchep{{\tt CalcHEP}}
\def\comphep{{\tt CompHEP}}
\def\lanhep{{\tt LanHEP}}
\def\slhaplus{{\tt SLHAplus}}

\def\suspect{{\tt SuSpect}}
\def\mbmb{m_b(m_b)}
\def\mt{m_t}
\def\dMb{\Delta m_b}
\def\dMq{\Delta m_q}
\def\delrho{\Delta\rho}
\def\bsgamma{b\to s\gamma}
\def\bsmu{B_s\to \mu^+\mu^-}
\def\gmuon{(g-2)_\mu}
\def\noi{\noindent}

%
%

\section{Introduction}

Nowadays there are two crucial problems in particle physics: the search for the 
 Higgs particle or more generally the unravelling of the mechanism of symmetry breaking and the nature of Dark Matter (DM). The
Higgs particle, responsible for symmetry breaking, is the cornerstone of the  Standard Model (SM) and of some
of its extensions. As long as the Higgs particle escapes detection there will be a missing link
in our understanding of the nature of fundamental interactions and the SM will be incomplete. 
As concerns the nature of DM we are facing a different issue: 
there is  robust  experimental evidence for DM, yet we have no additional evidence for the existence
of a stable massive particle which  can play the role of DM. For this  
we have  to consider extensions of the SM and assume an additional   discrete  unbroken   symmetry, 
 for example $Z_2$. This symmetry not only allows the lightest particle of the new physics model to be stable
 but it also usually makes this model conform more naturally with current data.

\begin{figure}[htb]
\label{RotCurv}
\centerline{
\includegraphics[width=6.5cm,angle=0]{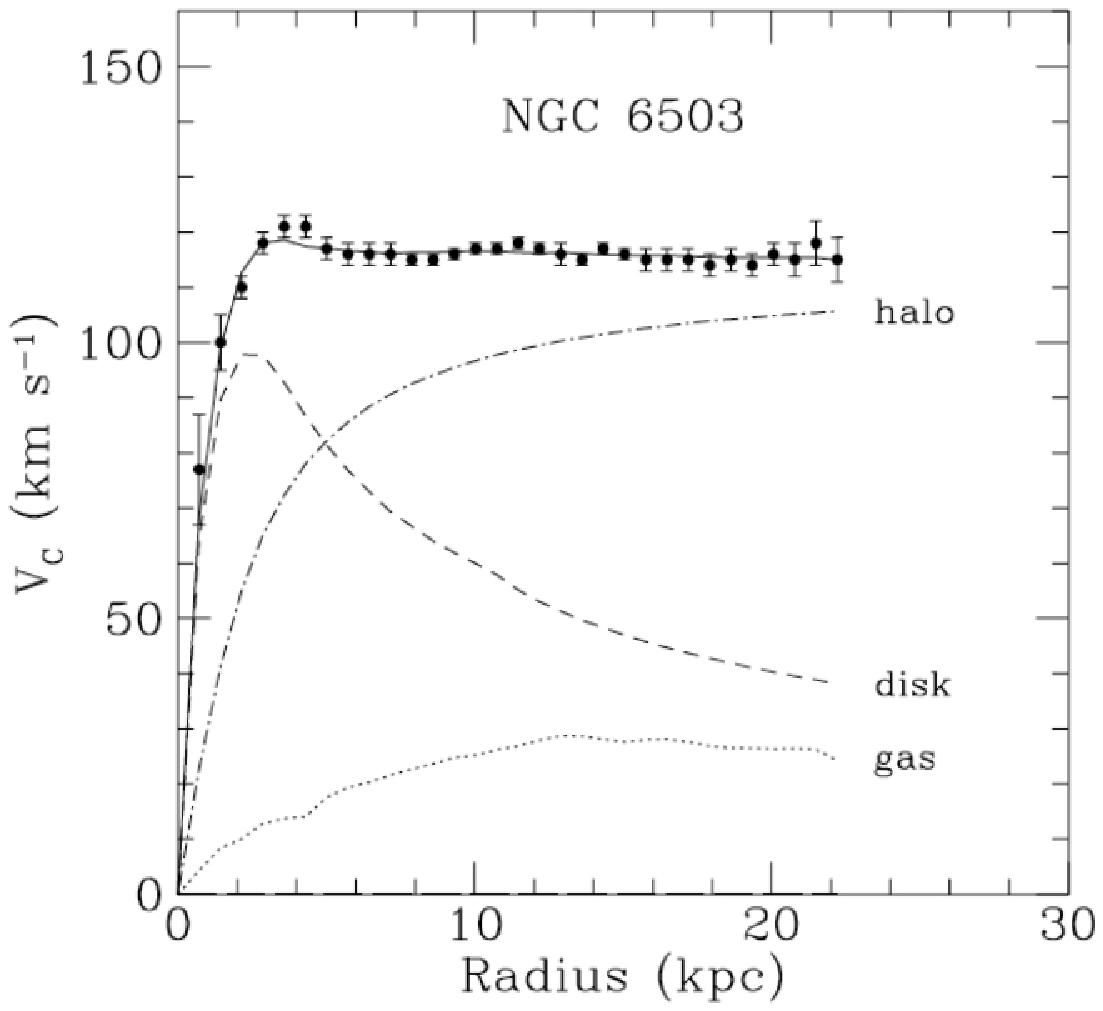} 
\includegraphics[width=7.8cm,angle=0]{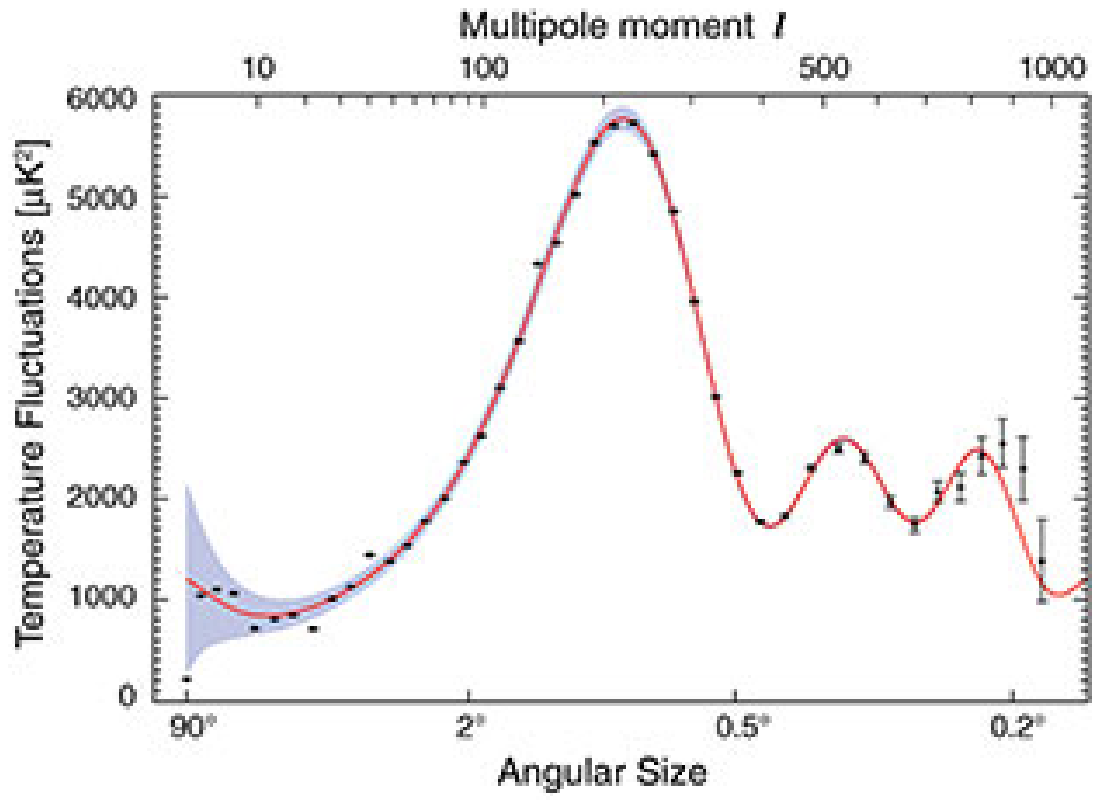}
           }
\caption{ a) Measured rotation  curve for the N6503 galaxy together with  
estimations of the contribution of the halo, disk and galactic gas to the rotation velocity 
(dashed). b) Temperature fluctuations measured by 
WMAP. The DM density is extracted from the second peak. }
\end{figure}

Let's review briefly the experimental  evidence for dark matter.  
First, the radial dependence of rotation curves of galaxies give strong evidence for DM.
Typical  rotation curves in spiral galaxies  show a plateau
at a distance of several kpc from the galactic center, see
Fig\ref{RotCurv}(a). The numerical value of the velocity 
at large distances is significantly larger than expected assuming only
visible matter. Furthermore such a plateau implies a  gravitational mass 
that increases linearly with the galactic radius, this does not corresponds to the 
distribution of visible matter. The rotation curve for the Milky Way allows to
estimate the density of DM in the  Sun orbit. This gives a value of $\rho_{DM}
\approx 0.3GeV/cm^3$ for the local density. This number enters the computation of the 
signals for DM direct and indirect detection that will be described below. 

Very precise estimates of the amount of DM were obtained from 
the  WMAP measurement  of  fluctuations in the  microwave
background temperature\cite{Bennett:2003bz,Spergel:2003cb}. 
Temperature   fluctuations, Fig\ref{RotCurv}(b), are connected to  fluctuations 
in the gravitational  potential in the  time of last scattering. 
Because ordinary  matter is in a plasma at high temperatures, 
 it can not generate such
fluctuations. 
Precise numerical analyses of WMAP results allow to extract the  total 
density of DM particles in the time of last scattering. 
Assuming  that the number of DM particles has not changed since then  we should have now a DM  density
around $1GeV/m^3$. More precisely, 
$$ \Omega h^2 = \frac{\rho_{DM}}{\rho_c} =  0.1099 \pm 0.0062$$ 
where $\rho_c$ is the critical density.
This amount of DM is in good agreement 
with  simulations of large structure formation in Universe. 
Indeed baryonic  matter itself  is not  able  to create galaxies because 
of the  fast expansion rate of the Universe.   
Because  DM particles have to be non-relativistic  
at the time of last scattering time, from WMAP measurements one can estimate its mass to be $M_{DM} > 10 KeV$.
This is sufficient to rule out neutrinos as the main component of DM.  
Therefore it is necessary to extend the SM to explain the nature of DM. 
The  precise measurement of $\Omega h^2$ provides a powerful  mean to discriminate various 
extensions of the SM that propose a DM candidate.

There are three kinds of  astroparticle experiments which allow in principle to detect DM particles
and measure some of their properties.
First, experiments for indirect DM detection such as 
                     PAMELA \cite{Adriani:2008zr, Adriani:2008zq},
HEAT \cite{Beatty:2004cy},   AMS01 \cite{Aguilar:2007yf},
Fermi \cite{Abdo:2009zk,Meurer:2009ir}, ATIC \cite{:2008zzr},
 HESS \cite{Aharonian:2009ah, Aharonian:2006au, Aharonian:2009nh} 
INTEGRAL \cite{Strong:2005zx}, Veritas\cite{Maier:2008vw}, 
EGRET\cite{Thompson:2008rw}
try to observe the products of  DM self annihilation in the galactic 
halo. The SM particles that are produced in this annihilation will decay to stable particles
including $\gamma$, $e^\pm$, $p(\bar{p})$ and neutrinos. 
Indirect detection experiments search primarily for $e^+$, $\bar{p}$, $\gamma$
as the  $e^-$ and $p$ channels suffer from a very large background and 
the neutrino signal is expected to be low.
Interpreting the results of indirect detection experiments requires a good understanding of both the background 
caused by galactic sources  as well as the
structure of galactic magnetic fields responsible for the   propagation of $e^+$ and  $\bar{p}$. 
For instance, the excess 
of positrons recently observed by PAMELA  can be caused either  by some exotic DM or some
galactic source like supernova.
Furthermore large uncertainties in the signal can be caused by a clumpy structure 
in the  DM  distribution (this  can increase the signal by a factor 20).

Direct detection experiments such as
Edelweiss\cite{Lemrani:2006ec}, DAMA\cite{Bernabei:2006mx},
 CDMS\cite{Akerib:2006jk,Ahmed:2009zw}, Xenon
\cite{Angle:2007uj,Aprile:2010um},
Zeplin\cite{Sumner:2005wv} or Cogent\cite{Aalseth:2010vx}
measure the recoil energy of the nuclei that would result from  an elastic  DM - nucleus collision  in a large detector.
To reduce the  cosmic rays background  such detectors are located deep underground.
We should  mention that DAMA has for several years found a   
positive result, such a signal has not yet been  confirmed by other experiments~\cite{Bernabei:2006mx}. 

High energy neutrinos  produced
as a result of  the annihilation of DM particles captured in the center of the Sun and the Earth,
are searched by   
Super-Kamiokande~\cite{Desai:2004pq}, Antares~\cite{Ernenwein:2008zz} or
IceCube~\cite{DeClercq:2008az}.
The rate of DM annihilation inside the Sun/Earth should be equal to the rate of DM capture
by the Sun/Earth. These  experiments are therefore similar to direct detection experiments  where the Sun or the Earth 
plays the role of the large detector. These experiments have not yet observed DM events. 

Dark matter can also be detected in accelerators such as the Tevatron or the LHC.
Despite the fact that the direct production of DM particles has a small cross section and that the DM particle
 escapes the detector  without leaving a track, a  DM particle could be detected at LHC.
 Indeed such particle appears in the decay chains of new particles that can be directly produced at a collider
 and its signature  is a large amount of missing energy~\cite{Baer:2008uu}.
It is therfore  possible that the LHC will soon shed light on the  two fundamental problems in particle physics: 
Higgs and dark matter.

\section{ Short review of theoretical models for DM}
Many extensions of the SM that can provide a  DM candidate have been proposed.
The best studied and most popular models among these are  supersymmertic 
models: the minimal supersymmetric model MSSM
\cite{Ellis:1983ew, Goldberg:1983nd} and its 
extensions such as the NMSSM \cite{Ellwanger:2009dp} and the CPVMSSM~\cite{Pilaftsis:1998pe}. 
In these models R-parity conservation guarantees the stability of the  DM particle. 
In models  with flat or warped extra dimensions  
\cite{Cheng:2002iz,Agashe:2004ci},  some parity that depend on the field in extra dimensions 
  is responsible for the stability of DM.
Furthermore there  are   models with extended gauge or
Higgs sectors \cite{McDonald:1993ex,Barger:2007nv} as well as little Higgs models~\cite{Hubisz:2004ft,Martin:2006ss,Belyaev:2006jh}. 
In these models the DM candidate can be either 
a Majorana fermion, a Dirac fermion, a vector boson or a scalar.

\section{Calculations needed for DM analyses.}
Because of the large number of  astroparticle   experiments   
and the large number of theoretical models we need software tools 
for the computation of DM properties  and  DM detection rates
in different  experiments. The 
general theoretical formulas for DM calculations are available  in the review
\cite{Jungman:1995df}. Detailed relic density  calculations in the MSSM can be found in 
\cite{Gelmini:1990je,Edsjo:1997bg} while direct detection formulas 
including  loop corrections and subleading terms was obtained in
\cite{Drees:1993bu}.  
The different  tasks which have to be solved are 
\begin{itemize}

\item{\it Calculation of DM relic density.} 
The formalism to calculate the DM density using the  freeze-out mechanism 
and based on  the DM annihilation cross sections was developed in \cite{Gelmini:1990je,Edsjo:1997bg}. 
One has to solve a differential equation that gives the temperature dependence of   DM   density.
The dependence on the underlying model  appears via the
calculation of the thermally  averaged cross section for DM annihilation.
A rough estimation  gives a value  for the annihilation 
cross section around $\sigma(\rm{v}) \rm{v} \approx 1pb\cdot c$. This corresponds 
to  a typical weak interaction cross section. Nevertheless   
agreement with  WMAP  results strongly constrain the parameters of the particle physics  model.
In addition to DM anihilation processes, processes involving other 
particles that are odd under the discrete symmetry and  whose masses are just above that of DM 
also contribute to the effective cross section since eventually all these particles will decay into
the DM and some other particles. 
The large number of processes involved and the fact that a priori the  matrix elements  needed
are not known means that relic density calculations in a generic model of
Dm can be challenging.

\item{\it Calculation of DM - nucleon cross sections.} These are required for the prediction of rates in direct 
detection  and neutrino telescope experiments. In the standard case 
one needs to compute the  DM-nuclei  scattering amplitude in the limit of small momentum transfer.
This is obtained from the 
DM-nucleon amplitude which is in  turn related to DM-quark amplitudes.

\item{\it Calculation of indirect detection signal}. 
In addition to the calculation of the DM annihilation cross section, the computation of the
spectra of photons, positrons and
antiprotons are required. The initial spectra can be easily obtained even  for a generic model.
For this one has to calculate all $2\to2$ annihilation cross sections and extract  the 
$\gamma$, $e^+$ and  $\bar{p}$ spectra using Pythia.
The $2\to2+\gamma$ processes also might have to be taken into
account\cite{Bringmann:2007nk}. 
The propagation  of $e^+$ and $\bar{p}$ can be done by 
solving the diffusion equation by the Green function method. A more 
precise treatment as well as the computation of the  background require
a Monte Carlo simulation.  

\item{\it Calculation of low energy  constraints.} 
Several experimental data  restrict the parameter space of SM extensions even though they are not directly related to DM
observables. 
These include precision measurements such as the muon anomalous magnetic moment, $g-2$,
or rare $B$-decays for example  $b\rightarrow s\gamma, B_s\rightarrow \mu^+\mu^-$. 
In most cases the theoretical prediction require the computation of higher order processes involving Feynman 
diagrams at the loop level. These are not completely automatized yet.
 
\item{\it LEP and Tevatron constraints}. 
High energy collider experiments are probing the SM and its extensions. 
Their results can be used to put constraints on the  Higgs 
mass, on new  channels in Z decay and on the mass of heavy exotic 
particles including the supersymmetric partners of SM particles.

\item{\it Calculation of LHC and ILC signals} 
The computation of signals associated with new particles produced at colliders
include the matrix element calculation for the production and for the decays of the new particles as well as  
Monte Carlo phase space integration and  cuts implementation.
Several tools have ben developed to perform these tasks. 
 \end{itemize}

\section{Review of software for DM calculation}

There are several  public codes used in DM calculations which were designed for the study of physics beyond the SM,
for a review of the different tools see~\cite{Skands:2005vi,Tools}.
Several codes perform the computation of the particle mass
spectrum in supersymmetric models, indeed large loop corrections are generic and need to be taken into account.
These codes also solve the renormalization group equations in supersymmetric scenarios 
with fundamental parameters defined at the GUT scale. Four codes were developed in the framework of the MSSM,
SoftSUSY~\cite{Allanach:2001kg},
Isajet~\cite{Baer:2004qq, Baer:2002fv} and SPheno~\cite{Porod:2003um},
while 
NMSSMTools~\cite{Ellwanger:2006rn} and CPsuperH\cite{Lee:2003nta}
deal with extensions of the MSSM. 
These codes also compute various low energy and collider constraints.
A special file interface SLHA~\cite{Skands:2003cj,Allanach:2008qq}  was designed 
for these programs. This interface facilitates their use in DM related code. 
The package HiggsBounds~\cite{Bechtle:2008jh} was designed for testing LEP and Tevatron accelerator constraints 
on the Higgs sector in generic models. Such constraints  are available in 
NMSSMTools  and Isajet but only for the specific  class of  models  they  support.

A very important tool for analysis of indirect detection experiments is the
GALPROP\cite{Strong:2009xj} program. It gives a numerical solution for 
the differential equation that describes the  propagation of different kind of particles 
in the galactic magnetic fields. Although this code is rather slow  it allows to take into 
account both DM signals and background galactic sources at the same time.

There are four public codes for  DM studies in supersymmetry,   
SuperIso\cite{Arbey:2009gu}, IsaTools \cite{Baer:2004qq, Baer:2002fv} 
DarkSUSY \cite{Gondolo:2004sc} and micrOMEGAs \cite{Belanger:2006is,Belanger:2008sj,  Belanger:2010gh}. 
All perform the computation of the DM relic density together with other observables that are
not necessarily related to DM.
SuperIso is a rather new code that is primarily dedicated to flavour physics in the MSSM, the SLHA is used for interfacing spectrum calculators.
IsaTools and DarkSUSY were also both developed for the MSSM. They calculate  the direct detection
and indirect detection rates as well as low energy and accelerator constraints.
IsaTools uses Isajet  to compute the particle spectrum while DarkSUSY also uses SuSpect or the SLHA.
DarkSUSY also calculates the neutrino rates from  DM annihilation in the
Sun and the Earth, furthermore DarkSUSY includes the propagation of cosmic rays. 
In particular it is interfaced with
GALPROP which allows to study both signal and background in  indirect detection
measurements. For MSSM applications, DarkSUSY is now the most complete
package. On the other hand IsaTools is based on Isajet, a tool for computation of 
signals for SM and its supersymmetric extensions at colliders and is therfore most suited for 
DM accelerator studies.

micrOMEGAs is the only  package for DM studies in generic
extensions of the SM model. Details of the techniques used 
in micrOMEGAs are  explained in the next section. micrOMEGAs  computes
the  DM relic density, direct detection and indirect detection 
rates. For the propagation of $e^+$ and $\bar{p}$, micrOMEGAs uses a  
Green function method which describes well the signals from DM annihilation  but 
does not allow to calculate the background. The neutrino rate from capture in celestial bodies  is not 
yet implemented in micrOMEGAs. Low energy and collider constraints are provided for some models and the predictions of collider
signals are obtained from CalCHEP which is included in micrOMEGAs.  The current version of  micrOMEGAs
contains the MSSM, NMSSM, CPVMSSM, the Little Higgs model~\cite{Belyaev:2006jh}, and a Dirac Neutrino
DM model~\cite{Belanger:2007dx}. 
   
Comparisons of IsaTools/DarkSUSY/micrOMEGAs showed  good agreement between the codes. 
In fact such cross checks  were used to remove several bugs in these packages.

\section{ Applications of   automatic matrix element calculators 
for dark matter studies.}

In principle it  is not necessary  to use only one universal program  to study  DM properties in any
model. On the other hand once such a tool has been tested and debugged for one specific extension of the SM, it can
rapidly and  straightforwardly be used for other models as well. An automatic approach therfore increases the 
reliability of the software and considerably reduces the time needed for developing new software as well
as the time required for the user to become familiar with a new package. 

 As mentionned above the most important computer task needed for DM studies is the computation of
matrix elements of various reactions which occur  in some specific model
of particle physics. In the last years several
automatic  calculators of matrix elements were developed: 
CompHEP \cite{Boos:2004kh}, CalcHEP\cite{Pukhov:2004ca},  FeynArts/FormCalc
\cite{Hahn:2000kx,Hahn:1998yk,Hahn:2006ig}, MadGraph\cite{Alwall:2007st,Maltoni:2002qb},
 Sherpa\cite{Gleisberg:2003xi}, and Omega \cite{Moretti:2001zz}.
In principle any of these could be used for DM related calculations in a  
generic model. 
Currently the idea of automatic matrix element generation for
DM observables  in a generic model  is realized in full scope only in the micrOMEGAs
package. This approach was first applied for the computation of the relic density~\cite{Belanger:2006is}.
In~\cite{Belanger:2008sj} a numerical algorithm for the calculation of the 
spin-dependent and spin-independent DM-nucleon amplitudes relevant for direct detection
 was proposed and implemented. This algorithm  which can be applied to a generic model
replaces the  usual symbolic computation of amplitudes  by means of Fiertz identities.
Recently in~\cite{Belanger:2010gh} an automatic approach for calculating the spectra of DM self annihilation in the galaxy
was designed and takes into account  processes with additional photon radiation  \cite{Bringmann:2007nk}.

The key point in micrOMEGAs' approach to  DM calculations  is the generation of shared libraries with matrix element codes.
The calculation of all matrix elements that enter
a relic density calculation  is computer time consuming and requires a lot of disk space
\cite{Gondolo:2004sc}. However for any particular set of model parameters in general only a
small number of annihilation channels are needed. micrOMEGAs therefore generates the code
 only for the channels as they are needed,
links them dynamically and stores them on the disk for subsequent usage.

Note that the idea of using automatic calculators in DM codes was also realised in 
 IsaTools and SuperIso albeit only in the context of the MSSM. In IsaTools, CompHEP  was used for
generating (co-)annihilation cross sections while SuperIso relies on
FeynCalc to  evaluate the cross sections. In principle both these codes  can be
generalized for  other models.

\section{Conclusion}

Several tools for the calculation of DM properties  and DM signals for  current and  future experiments are now available. 
The currently most developed codes are DarkSUSY and micrOMEGAs. The existence
of several independent codes  is very important for cross checking the results and
for understanding uncertainties  which result from   different 
technical implementation of the same algorithms. 
There are several auxiliary  tools  designed for the computation  of the particle spectra and
couplings  as well as for calculation of low energy and high energy constraints.
The development of interface
protocols  for  data exchange between  such programs  is needed.

\section{Acknowledgments}
     This work was supported in part by the GDRI-ACPP of CNRS, by the ANR
project ToolsDMColl, BLAN07-2-194882, by the
Russian foundation for Basic Research, RFBR-08-02-92499-a, RPBR-10-02-01443-a and by a State contract No.02.740.11.0244. 
The visit of A.P. to Jaipur was funded  by the organizing 
committee  and by the grant RFBR-10-07-08004-z.

\end{document}